\documentclass[12pt]{article}
\usepackage{graphicx} 
\usepackage [english] {babel}
\usepackage[small]{titlesec} 
\usepackage[utf8]{inputenc} 
\usepackage[T1]{fontenc}
\usepackage[english]{babel}
\usepackage[left=2cm,right=2cm,top=2cm,bottom=2cm,bindingoffset=0cm]{geometry}

\graphicspath{{figures/}}

\usepackage{setspace}
\usepackage{subcaption}
\usepackage{youngtab}
\usepackage{epigraph}

\usepackage[labelfont=bf, margin=20pt,font={small}]{caption}

\usepackage{amsmath}
\usepackage{esint}
\usepackage{bm}
\usepackage{
	tikz,
	amsmath,
	amssymb,
	amsfonts,
	amsthm,
	amscd,
	comment,
	epsfig,
	color,
	setspace,
	bbold,
	verbatim,
	slashed
}
\usepackage{upgreek}
\usepackage{tikz}
\usetikzlibrary{arrows}
\usepackage{mathrsfs}
\usepackage{multirow}
\usepackage{simpler-wick}
\usepackage{cite}

\newcommand\beq{\begin{equation}}
	\newcommand\eeq{\end{equation}}

\tikzset{cross/.style={cross out, draw=black, minimum size=2*(#1-\pgflinewidth), inner sep=0pt, outer sep=0pt},
	cross/.default={5pt}}
\usepackage{mathtools}
\usepackage{graphicx}
\usepackage{authblk}
\usepackage{xcolor}
\usepackage{bbold}
\usepackage[makeroom]{cancel}
\usepackage[hidelinks]{hyperref}

\definecolor{linkcolor}{rgb}{0,0,1} 
\definecolor{urlcolor}{rgb}{0,0,1} 

\hypersetup{pdfstartview=FitH, linkcolor=linkcolor,urlcolor=urlcolor,citecolor=urlcolor, colorlinks=true}

\title{\textcolor{black}{Surprises of the Klein paradox}}

\author[1]{T.~R.~Ahmadov\thanks{\href{mailto:akhmedov.tofik@bk.ru}{akhmedov.tofik@bk.ru}}}
\author[2,3]{E.~T.~Akhmedov\thanks{\href{mailto:akhmedov@itep.ru}{akhmedov@itep.ru}}}
\author[2]{V.~I.~Lapushkin\thanks{\href{mailto:volapushkin@gmail.com}{volapushkin@gmail.com}}}
\affil[1]{\itshape Azerbaijan State Oil and Industry University, Baku, Azerbaijan}
\affil[2]{\itshape Institutskii per, 9, Moscow Institute of Physics and Technology, 141700, Dolgoprudny, Russia}
\affil[3]{\itshape Academician Kurchatov Square, 1, NRC ''Kurchatov Institute'', 123182, Moscow, Russia}
\date{\today}

\begin{document}
\maketitle

\begin{abstract}
We consider second quantized $1+1$ dimensional Dirac fermions with mass $m$ in background electric fields. This theory mimics the following situations. We consider two parallel capacitors with opposite directions of the electric field $E$ in them. The potential energy difference in one of them is equal to $V_0 > 2m$ and in the other to $V+V_0$, with $0<V<2m$. According to general physical understanding of the Schwinger effect, each capacitor separately will create pairs and there will be oppositely directed currents, which will not be equal to each other. The total current will not vanish. In this paper we show that this is too naive an observation: the limiting expectation value of the current operator, as $t\to +\infty$, is actually zero, if the above form of the potential with $0<V<2m$ is instantaneously turned on. The non-vanishing current is obtained only if $V>2m$. We also consider the same theory in the presence of the infinite wall and a deep well $V_0>2m$. We show that in such a potential the limiting expectation value of the current is zero in the Gaussian approximation. We discuss the relation of these observations to the Schwinger effet and to the decay of supercritical nuclei.  
\end{abstract}

\newpage

\tableofcontents

\newpage
    
\section{Introduction}

According to Schwinger \cite{Schwinger:1951nm}, if in a capacitor the electric field strength $E$, the distance between plates $L$, and the electric charge $e$ form a potential energy difference $V=eEL$ that is greater than twice the rest energy $V>2m$ (we set the speed of light and Planck's constant to unity) of the electron, there is a breakdown of the vacuum and creation of electron-positron pairs from the vacuum. Correspondingly, one will detect a current of the created pairs. A closely related phenomenon is due to Klein \cite{Klein} (see also \cite{Calogeracos:1999yp}), which can already be seen in $1+1$ dimensional electrodynamics. See the analysis of this phenomenon within many-particle \cite{Calogeracos:1998rf, Gavrilov:2015yha, Nikishov:1970br, Nikishov1985} and one-particle frameworks \cite{Nikishov:1970br, ternov2024paradoks15483}.

Consider now two parallel capacitors with opposite directions of the electric field in them (as schematically shown in Fig. \ref{Fig.1:In_modes_for_big_wall}). Let the potential energy difference in one of them be equal to $V_0 > 2m$ and in the other equal to $V+V_0$, with $0<V<2m$. Naively, according to \cite{Schwinger:1951nm}, each capacitor separately will create pairs. There will be oppositely directed currents, which will not be equal to each other. The total current will not vanish. In this paper we show that this is too naive an observation.

Namely, we show that the stationary current created in such a system is zero independently of the value of $V_0>0$, if $V<2m$. A non-zero current will appear only if $V>2m$, independently of the value of $V_0>0$. We do this in $1+1$ dimensional theory in the Gaussian approximation: when the background electric field is fixed and only the fermions without self-interactions are quantized.

Perhaps it is worth stressing that in real physical situations to realize the phenomenon that we observe the width of the well, $a$, and the size of the system should be comparable to the Compton wavelength $1/m$. Furthermore, of course, one has to check the situation in $3+1$ dimensions separately, which will be done elsewhere. But it is hard to expect that the presence of the tangential momenta of the modes will drastically change our observations. Furthermore, if the Schwinger phenomenon in the system that we consider appears only due to boundary effects and/or in the interacting theory (i.e. beyond the Gaussian approximation, when the electric field is also quantized), that should completely change one's intuition on the physics of such effects.

To explain our observations in simple terms, in the next section we give basic physical arguments with few equations and only after that we continue with technical details. We also show that in $1+1$ dimensions, in the presence of an infinite wall and a well of depth $V_0 > 2m$, there is no particle creation in the Gaussian approximation. The latter situation is relevant for the consideration of the supercritical nuclei.

There is another reason why we consider the phenomenon of particle creation within the framework of the Klein paradox. The phenomenon of particle creation in four-dimensional theories 
is very hard, at least  beyond the Gaussian approximation (see e.g. \cite{Akhmedov:2026wew,Akhmedov:2024npw,Akhmedov:2024rkt,Akhmedov:2024lce,Akhmedov:2023zfy} and \cite{Akhmedov:2021rhq} for a review). Hence, one needs an analog of something like the Ising model, which serves for a basic understanding of phenomena in quantum field theory and critical phenomena, to better understand the phenomenon of particle creation in simple terms. The Klein phenomenon seems to be the model of that kind.
    
\section{Explanation of the phenomenon in simple terms}

In this section we describe the phenomenon under consideration in simple terms and with little calculations. 
We consider a quantum Dirac field in $1+1$ dimensions in a background electric potential:

\begin{eqnarray}
\label{deq1}
    \Big(i\gamma^{\mu}\partial_{\mu} + e \gamma^\mu A_\mu -m\Big)\hat{\psi}(t,x)=0, \\ A_\mu = (A_0, 0), \;\;
    \gamma^{0} = \begin{pmatrix}
	1 & 0\\
	0 & -1
	\end{pmatrix}, \;\; \gamma^{1} = \begin{pmatrix}
	0 & i\\
	i & 0
	\end{pmatrix}.    \nonumber 
\end{eqnarray}
The background field is classical and not dynamical, i.e. fixed by external sources. Our goal is to calculate the expectation value of the fermion current operator $\hat{\psi}^{\dagger}(t,x)\hat{\alpha}\hat{\psi}(t,x)$, $\hat{\alpha} = \gamma^0 \gamma^1$ 
for the second-quantized fermion theory as $t \to \infty$ (in proper units set by $m$ and parameters of $A_0$) if the potential $A_0(x)$ is rapidly turned on (again in the same proper units) at some moment in time. We assume that the initial state is the standard Minkowski (Poincare invariant) vacuum of the fermion field, i.e. for the case $A_0 = 0$. Using the intuition gained in the previous paper \cite{Akhmedov:2025jtk}, we can find the current as $t\to \infty$ in such a time-dependent situation by calculating it for certain states in a stationary (eternally existing) background field $A_0(x)$, which is a limiting form of $A_0(t,x)$ as $t\to \infty$. Let us explain in simple terms how this can be done and what it has to do with the Klein paradox. 

To calculate the expectation value of the current, one has to find the spectrum of modes for the Dirac field, which follows from eq. (\ref{deq1}). Taking into account that we consider stationary but spatially inhomogeneous background fields $A_0$, we perform the Fourier transformation of the time variable $t$ and look for exact spatial parts of the modes in the background fields. 

Let us start with the situation when there is no background field, $A_0 = 0$. In such a case, the spectrum of energies $\epsilon$ consists of two parts: $\epsilon\in(-\infty,-m)\cup (m,+\infty)$. The region $\epsilon \in (-m,m)$ is the forbidden zone. Furthermore for each allowed value of $\epsilon$ there are two solutions: the left-moving modes:
\begin{gather}
\label{leftfree}
    \psi_{\epsilon L}(x) = \frac{1}{\sqrt{2k_\epsilon\left|\epsilon-m\right|}}
    \begin{pmatrix}
        -ik_\epsilon \\
        \epsilon - m
    \end{pmatrix}e^{-ik_\epsilon x} 
\end{gather}
and the right-moving ones:
\begin{gather}
\label{rightfree}
    \psi_{\epsilon R}(x) = \frac{1}{\sqrt{2k_\epsilon\left|\epsilon-m\right|}}
    \begin{pmatrix}
        ik_\epsilon \\
        \epsilon - m
    \end{pmatrix}e^{ik_\epsilon x} ,
\end{gather}
where $k_\epsilon = \sqrt{\epsilon^2-m^2}$.  

Note that under complex conjugation one maps the $\epsilon\in (m,+\infty)$ part of the spectrum into the one with $\epsilon\in(-\infty,-m)$, and the $L$ and $R$ modes are exchanged with each other. Each mode contributes the expression $\psi_{\epsilon L,R}^{\dagger}(x)\hat{\alpha}\psi_{\epsilon L,R}(x)$ to the expectation value of the current and the contribution of each mode (left and right movers) separately is not zero. At the same time, while for the positive energy modes, $\epsilon > m$, the contribution to the current of each $L$ and $R$ modes is directed along the momentum $k_\epsilon$, for the negative energy modes, $\epsilon < - m$, this contribution to the current has the opposite direction with respect to the momentum $k_\epsilon$. These observations are important for the Klein phenomenon and for more complicated situations, which are considered below.
 
For each value of the energy $\epsilon \in (-\infty,-m)\cup (m,+\infty)$ separately, the left-moving modes cancel the right-moving ones in the total expression for the expectation value of the current $\left\langle st\left|\hat{\psi}^{\dagger}(t,x)\hat{\alpha}\hat{\psi}(t,x)\right|st\right\rangle$. This is the story for the standard Minkowski vacuum state $|st\rangle = |0\rangle$ when $A_0 = 0$. Of course even if $A_0 = 0$ one can always organize such a state for which the total current is not zero: everything depends on which levels are filled in and which ones are empty. 

As usual one can understand the Minwkowski vacuum state under consideration in several different ways. One way to understand it is as the standard Fock space ground state, in which there are no negative energy levels for both one-dimensional electrons and positrons. Another way --- as the Dirac's sea, which is filled by electrons up to some energy $E$. The latter usually is chosen to be equal to $E=-m$, but actually the value of the expectation value of the current will be zero even for such a state, in which all levels for both $L$ and $R$ modes up to any $E$ are filled in.


\begin{figure}[t]
    \centering
    \includegraphics[width=0.65\linewidth]{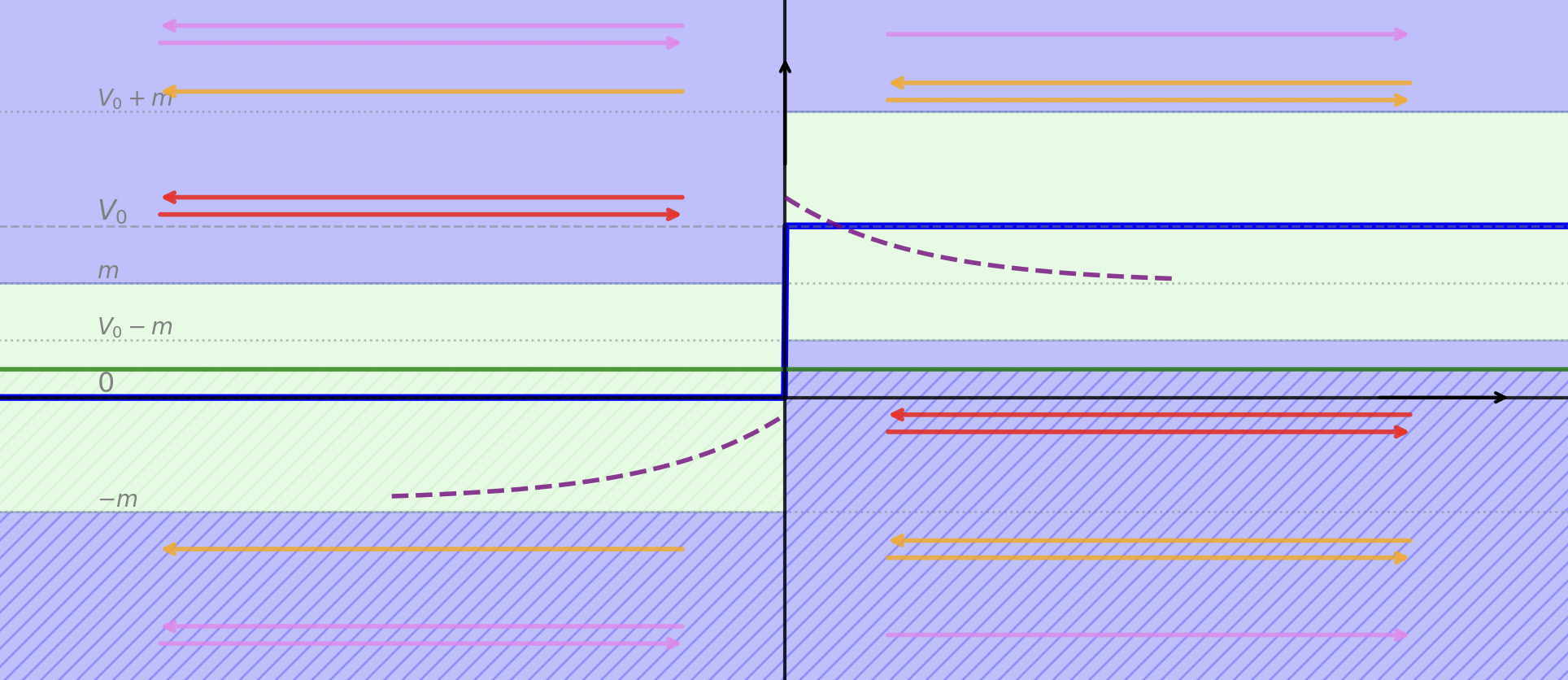}
    \caption{Illustration of modes in the presence of the step potential. For $\epsilon \in (-\infty, - m) \cup (V_0 + m, +\infty)$ the $R$ modes are linear combinations of the falling on the barrier and reflected waves at $x<0$, and are single transmitted waves for $x>0$. For the $L$ modes the situations in $x<0$ and  $x>0$ are exchanged with each other. At the same time, for $\epsilon \in (-m, V_0 + m)$ the modes are linear combinations of left- and right-moving waves that exponentially decay into forbidden zones either towards $x<0$, for $\epsilon \in (-m, V_0-m)$, or towards $x>0$, for $\epsilon \in (m,V_0+m)$.}
    \label{Fig:In_modes_plot1}
\end{figure}

\begin{figure}[t]
    \centering
    \includegraphics[width=0.65\linewidth]{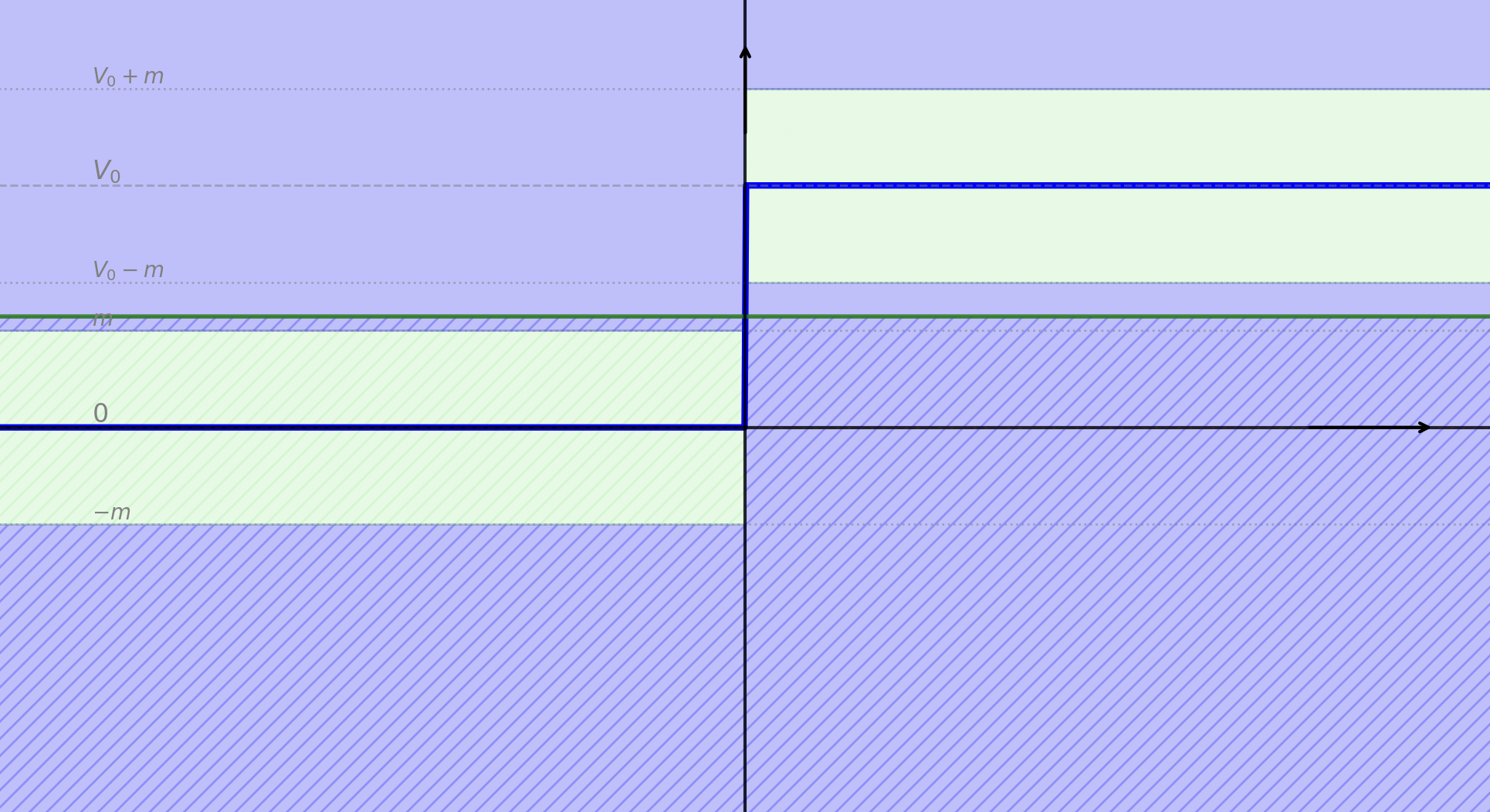}
    \caption{}
    \label{Fig:In_modes_plot2}
\end{figure}

Now, if there is a background potential of the form $e A_0(x) = V_0 \, \theta (x)$, where $\theta(x)$ is the Heaviside step function (see Fig. \ref{Fig:In_modes_plot1} or \ref{Fig:In_modes_plot2}), the properties of the spectrum change. Meanwhile the modes for this concrete situation can be found in \cite{Akhmedov:2025jtk}, we discuss similar modes for a different form of $A_0(x)$ in detail below in the main body of the paper. When $V_0 < 2m$ the changes in the spectrum and in the expectation value of the current (for the simplest states) are minor. In fact, for $\epsilon \in (-\infty, - m) \cup (V_0 + m, +\infty)$ we still have a doubly degenerate continuous spectrum where left-moving modes are present simultaneously with right-moving ones: the $R$ modes are linear combinations of the falling on the barrier and reflected waves at $x<0$, and are single transmitted waves for $x>0$. For the $L$ modes the situations in $x<0$ and $x>0$ regions are exchanged with each other.

At the same time, for $\epsilon \in (-m, V_0 - m) \cup (m, V_0 + m)$ the spectrum is continuous but not degenerate. For this range of energies, the modes are linear combinations of left- and right-moving waves that exponentially decay into forbidden zones either towards $x<0$, for $\epsilon \in (-m, V_0-m)$, or towards $x>0$, for $\epsilon \in (m,V_0+m)$. 

As a result, for certain type of natural states (which actually do not respect Poincare symmetry) the contribution to the expectation value of the current from each value of energy $\epsilon$ is zero: while for $\epsilon \in (-\infty, - m) \cup (V_0 + m, +\infty)$ the left movers cancel right movers, for $\epsilon \in (-m, V_0 + m)$ each mode separately contributes zero independently of the type of the state. The type of states that we can consider are as follows: we fill in by the Dirac sea all the levels below some value $E$, which can be taken arbitrarily. The surface of the Dirac sea is illustrated by the green line on Fig. \ref{Fig:In_modes_plot1}.

At the same time, when $V_0 > 2m$, as is shown on Fig. \ref{Fig:In_modes_plot2}, one encounters a completely new situation in the spectrum. On top of the modes described in the previous paragraph\footnote{Note that now a non-degenerate spectrum of the modes exponentially decaying into forbidden zones is present for $\epsilon \in (-m, m) \cup (V_0-m,V_0+m)$.}, we have modes in the so-called Klein zone $\epsilon \in (m, V_0-m)$. In this zone, the left-moving waves with $\epsilon > m$, which are defined for $x<0$, are glued to the right-moving waves with $\epsilon < V_0-m$, which are defined for $x>0$, and vice versa (see \cite{Akhmedov:2025jtk} for the details). That is necessary to have the current conservation for each mode separately: Note that waves below the forbidden zone $\epsilon < V_0-m$ for $x>0$ have the inverse direction of the current $\psi_\epsilon^{\dagger}(x)\hat{\alpha}\psi_\epsilon(x)$ as compared to the modes above the forbidden zone\footnote{Actually, such a phenomenon in the Klein zone allows for the presence of transition coefficients greater than one for the single particle problem, which is the essence of the Klein paradox in the single particle problem.} $\epsilon > m$, for $x<0$. The gluing of the left modes to the right ones in the Klein zone in its own right is necessary to obtain canonical commutation relations for the ladder operators and for fermion fields: The commutation relations can be connected to the current conservation for each mode separately.  

In the case under consideration for the static situation, when the potential $A_0(x)$ is always present from the past to future infinity, one can organize different Fock space ground states. In fact, in the situation in question there is no natural way to assign notion of particles and antiparticles to the modes and defining negative and positive energy states \cite{Akhmedov:2025jtk}. 
For example, one can consider such a state that all levels with $\epsilon < E$ are filled (see the green line and shaded region on Fig. \ref{Fig:In_modes_plot2}). Natural values of $E$ include $E=-m$, $E=V_0-m$ or $E = (V_0 - 2m)/2$, but $E$ can actually take any value without any effect on the total current: For a state of such a type the expectation value of the current $\left\langle st\left|\hat{\psi}^{\dagger}(t,x)\hat{\alpha}\hat{\psi}(t,x)\right|st\right\rangle$ is zero, because every left moving mode cancels the right moving one in the total expression of the current (see \cite{Akhmedov:2025jtk} for the details), while modes penetrating the forbidden zones contribute zero.

However, there is another possible state in the case under consideration. Let all the levels up to $E=-m$ are filled by left and right movers (i.e. by the standard Dirac sea), but the left moving levels with $V_0 - m > E > m$ are filled by electrons, while all the right moving levels are empty. The same state can be viewed in a different way: all left moving levels with $(V_0 - m)/2 > E > m$ are filled by electrons, while all the right moving levels with $V_0 - m > E > (V_0 - m)/2$ are filled by positrons\footnote{Recall that in the Klein zone there is a mixture between left moving ``electron'' modes and right moving ``positron'' modes and vise versa. Hence, how to interpret the situation depends on the way we define the zero energy level in the static situation.}. In any case, for such a state the expectation value of the current is obviously not zero, because contributions to it from the modes in the Klein zone are not compensated between each other \cite{Akhmedov:2025jtk}. Rather than that we have the situation in which left moving electrons add up with right moving positrons.

It can be shown that the state described in the last paragraph is realized as the final out state (i.e. as $t\to +\infty$) in the situation when one rapidly turns on the background field at some moment of time \cite{Akhmedov:2025jtk}, i.e. when $e A_0(t,x) = V_0 \, \theta (x) \, \theta(t)$. Vague explanation of the phenomenon is as follows: all the levels below the forbidden zone are filled in by the Dirac sea of electrons before the moment of turning on the external potential. That is our choice of the initial state. Then, if one instantaneously shifts up or down the half ($x>0$) of the forbidden zone, the levels below the forbidden zone remain filled in. Of course this is a very vague picture based on the intuition gained from solid state physics for non-relativistic electrons and one should take this explanation very carefully and check the situation by a calculation in non-stationary case along the lines of \cite{Akhmedov:2025jtk}.

The resulting expectation value of the current does not depend neither on time nor on spatial coordinate and can be related to the transition coefficients for the modes in the Klein zone \cite{Akhmedov:2025jtk}. At intermediate values of time the expectation value of the current is not stationary and is spatially inhomogeneous. As we explained, the state that we are discussing here is the final stationary state in the problem with the time-dependent background field $A_0(t,x)$, when the second quantized fermion theory is Gaussian.

Furthermore, the reason why the state under consideration is stationary (the resulting current is time independent) has to do with the fact that we have a continuous number of states, which are filled in by electrons, and which have to be filled in by the particles flowing in the current. We expect that if one will consider the system in a box and on a lattice (i.e. the field theory in question will have UV and IR regularization and, hence, the number of states in the system will be finite) and add interactions (i.e. one will go beyond the Gaussian approximation), then our state will decay into the one in which all the levels below $E = (V_0-m)/2$ for both $L$ and $R$ modes will be filled by electrons. This can be predicted on general physical grounds. Thus, we expect that in real physical situation there can be several different time scales.

This concludes the explanation in simple terms of the results of our previous paper \cite{Akhmedov:2025jtk}. We hope that this will help to better and deeper understand our observations below, which contain rather tedious calculations.
    
\section{The basis of modes for the case of the wall and a well}

\begin{figure}[h!]
		\centering
        \includegraphics[width=0.7\linewidth]{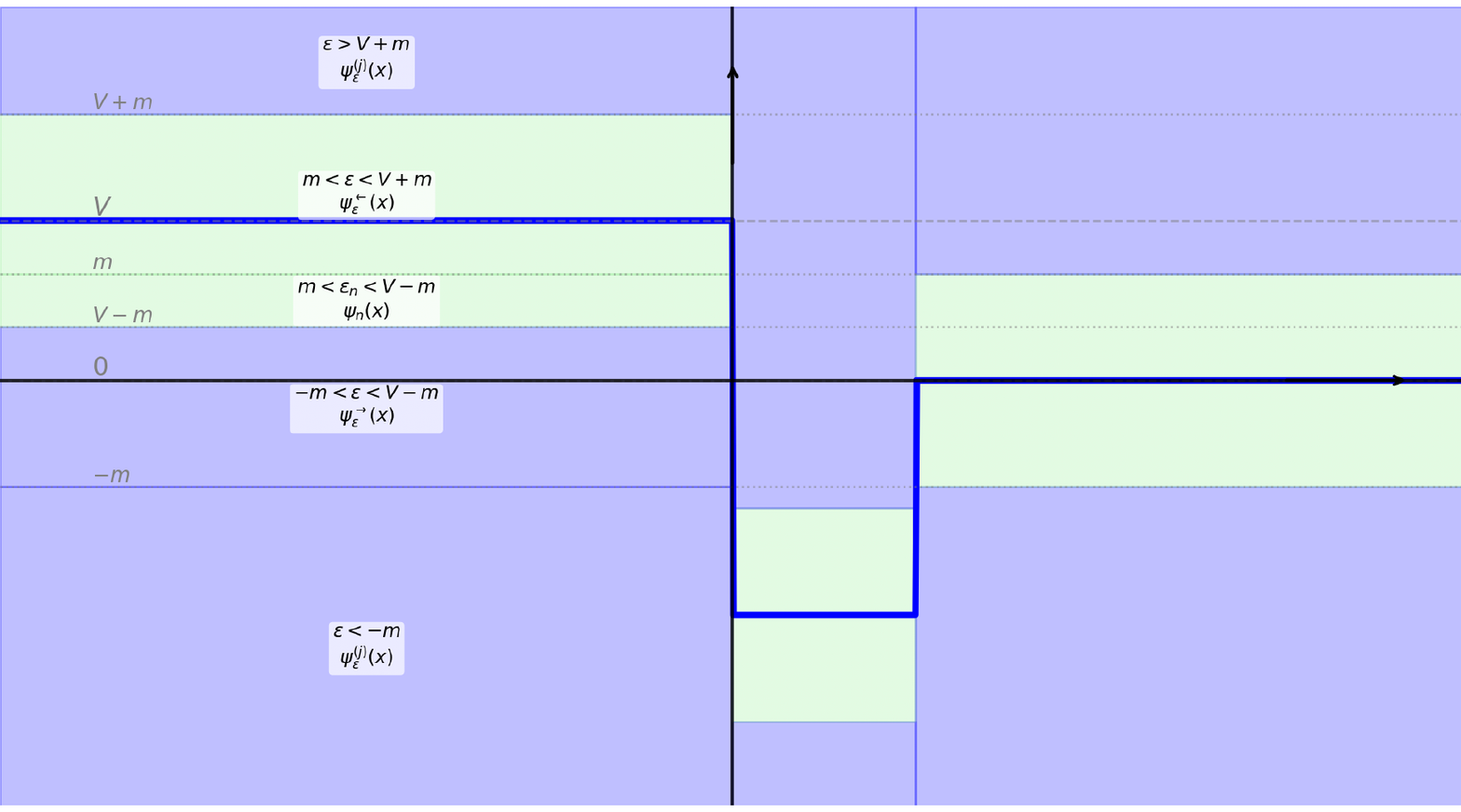}
		\includegraphics[width=0.7\linewidth]{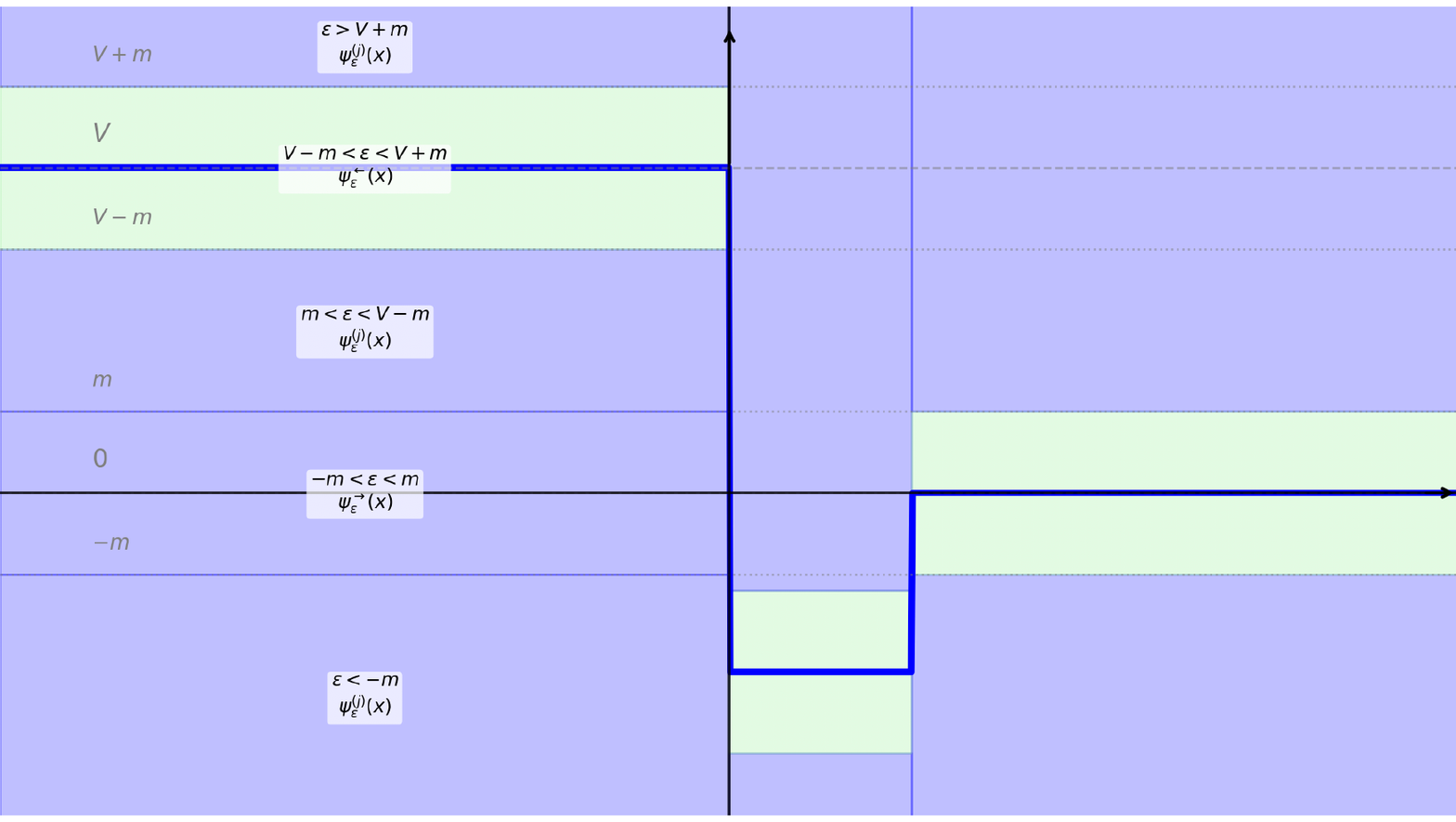}
        \caption{In the upper picture the height of the left wall above zero $V$ is smaller than $2m$, while in the lower picture it is greater than $2m$. In both cases the depth of the well $V_0$ is greater than $2m$. We show that if such a potential is switched on rapidly at some moment in time, a nonzero current appears as $t\to \infty$ only in the lower picture.}
		\label{Fig.1:In_modes_for_big_wall}
\end{figure}

We consider the situation depicted in Fig. \ref{Fig.1:In_modes_for_big_wall}, i.e. the background electric potential is:
	\beq\label{pot}
    -eA_0(t,x)=\theta(-x)V-\theta(x)\theta(a-x)V_0.
    \eeq
Such a potential can be described in two different ways: either as the presence of two capacitors with the potential energy differences $V_0$ and $V+V_0$ or as the wall of hight $V$ above zero and a well of depth $V_0$.

We show that independently of the depth of the well $V_0$ (even if $V_0 > 2m$) particle creation happens {\it only} when the height of the wall $V$ is greater than $2m$. Namely, we show that if one starts with $A_0=0$ and the Minkowski vacuum state, and then at $t=0$ rapidly turns on the potential (\ref{pot}), the stationary value of the current as $t\to +\infty$ vanishes if $V< 2m$. This holds true even if $V_0 > 2m$. This fact is proved as a theorem for a generic form of the potential in section 4 of \cite{Akhmedov:2025jtk}, but here we want to redo the explicit calculation for a simple concrete situation, give a simple explanation of the effect and stress the importance of these observations for the Schwinger effect and for the decay of supercritical nuclei. 

In terms of the discussion in the previous section, the final stationary state in the potential under consideration corresponds to the Dirac sea filling all the levels up to some energy level $\epsilon = E$ less than $V < 2m$. All the modes above $\epsilon=-m$ are exponentially decaying in the forbidden zones and contribute zero to the expectation value of the current because they are linear combinations of left- and right-moving waves. At the same time, the left- and right-moving modes from the seemingly present Klein zone $-V_0+m < \epsilon < -m$ are actually completely filled (both left and right moving sectors) and cancel each other in the total contribution to the expectation value of the current. This happens due to the simultaneous presence of the left and right walls of the potential under consideration. 

Moreover, the expectation value of the current will not be zero as $t\to +\infty$ only if $V>2m$, because in this case there is a Klein zone for $m < \epsilon < V - m$, and in it only the left-moving modes for $m < \epsilon < (V - m)$ are filled by electrons. This is the simple explanation, in the language of the previous section, of the long but straightforward calculation presented below in this and the next sections.

Let us explain why we find these observations rather surprising. In fact, each wall in the potential $A_0$ plays the role of a narrow capacitor with the potential energy difference between its plates equal to $V+V_0$ (for the left wall) and $V_0$ (for the right wall), respectively. Naively, the current created by each capacitor should be proportional to the potential energy difference if it is greater than $2m$. But these currents are directed in opposite directions. Hence, naively when $V_0> 2m$ these currents can compensate each other only when $V=0$. In the next section we show that these observations are far too naive. 

In this section we construct the basis of modes for the potential under consideration.
The Dirac Hamiltonian of the system is\footnote{Following the logic of the previous section we consider the stationary situation with the eternally existing background potential.}:

\begin{equation}
    \hat{H} = -i\hat{\alpha}\partial_{x}+m\hat{\beta}+\theta(-x)V-\theta(x)\theta(a-x)V_0,
\end{equation}
where $\hat{\alpha}$ and $\hat{\beta} = \gamma_0$ are the Dirac alpha and beta matrices. 

In this section we start with the definition of the spectrum of this Hamiltonian for the situation with a tall left wall, $V>2m$. The modes for the case $V<2m$ will be described in the next section. We always assume that $V_0 > 2m$. To find the spectrum of modes we have to solve the linear system of differential equations of first order for two functions. Hence, there are two linearly independent solutions for energies $\epsilon\in(-\infty;-m)\cup(m,V-m)\cup(V+m,+\infty)$, essentially corresponding to the left- and right-moving waves. We will denote them as $\psi^{(1)}_{\epsilon}(x)$ and $\psi^{(2)}_{\epsilon}(x)$. At the same time, for energies $\epsilon\in(-m;m)$ the modes are exponentially decaying as $x\rightarrow+\infty$. We will denote these modes as $\psi^{\rightarrow}_{\epsilon}(x)$. Similarly, we denote the modes decaying at $x \to - \infty$ as $\psi^{\leftarrow}_{\epsilon}(x)$ for $\epsilon\in(V-m;V+m)$ (Fig. \ref{Fig.1:In_modes_for_big_wall}). We refer to the latter two as the modes from the forbidden zones.
	
We need to know the explicit form of only $\psi^{(1)}_{\epsilon}(x)$ and $\psi^{(2)}_{\epsilon}(x)$, because the current for the decaying modes from the forbidden zones is zero. In fact, the expressions $\psi^*_{\epsilon,\leftarrow}(x)\hat{\alpha}\psi_{\epsilon,\leftarrow}(x)$ and $\psi^*_{\epsilon,\rightarrow}(x)\hat{\alpha}\psi_{\epsilon,\rightarrow}(x)$ are independent of $x$ due to current conservation. The latter constants are zero for the modes from the forbidden zones because they decay at one of the spatial infinities.

Thus, the modes that are of interest to us, which can provide non-zero contribution to the total current, have the following form:

\begin{equation}
	\psi^{(1)}_{\epsilon}(x) = \left[i \, e^{- i \,  {\rm arg}[G_2(\epsilon)]}\cos(\theta_{\epsilon})-\frac{G^*_2(\epsilon)\sin(\theta_{\epsilon})}{\sqrt{G_1^2(\epsilon)-|G_2(\epsilon)|^2}}\right] \, \frac{\psi_{\epsilon}(x)}{\sqrt{G_1(\epsilon)}} + \sin(\theta_{\epsilon})\sqrt{\frac{G_1(\epsilon)}{G_1^2(\epsilon)-|G_2(\epsilon)|^2}} \, \psi^*_{\epsilon}(x),
\end{equation}
and

\begin{equation}
	\psi^{(2)}_{\epsilon}(x) = \left[i e^{-i {\rm arg}[G_2(\epsilon)]}\sin(\theta_{\epsilon})+\frac{G^*_2(\epsilon)\cos(\theta_{\epsilon})}{\sqrt{G_1^2(\epsilon)-|G_2(\epsilon)|^2}}\right] \, \frac{\psi_{\epsilon}(x)}{\sqrt{G_1(\epsilon)}}-\cos(\theta_{\epsilon})\sqrt{\frac{G_1(\epsilon)}{G_1^2(\epsilon)-|G_2(\epsilon)|^2}} \, \psi^*_{\epsilon}(x).
\end{equation}
Here $\psi_{\epsilon}(x)$ is defined piecewise (here $a$ is the width of the well):

\begin{gather}
    \psi_{\epsilon}(x)=\begin{pmatrix}
		i\sqrt{q_{V,+}q_{V,-}}\\
		q_{V,-}
	\end{pmatrix}e^{i\sqrt{q_{V,+}q_{V,-}} \, x}, \quad \text{for} \quad x<0; \nonumber \\
    \psi_{\epsilon}(x) = \frac{1}{2} \, \left[\frac{q_{V,-}}{q_{V_0,-}}+\sqrt{\frac{q_{V,+}q_{V,-}}{q_{V_0,+}q_{V_0,-}}}\right] \, \begin{pmatrix}
		i\sqrt{q_{V_0,+}q_{V_0,-}}\\
		q_{V_0,-}
	\end{pmatrix}e^{i\sqrt{q_{V_0,+}q_{V_0,-}}x}+\nonumber \\
	+ \frac{1}{2} \, \left[\frac{q_{V,-}}{q_{V_0,-}}-\sqrt{\frac{q_{V,+}q_{V,-}}{q_{V_0,+}q_{V_0,-}}}\right]\begin{pmatrix}
		-i\sqrt{q_{V_0,+}q_{V_0,-}}\\
		q_{V_0,-}
	\end{pmatrix}e^{-i\sqrt{q_{V_0,+}q_{V_0,-}}x}, \quad \text{for } \quad 0<x<a; \\
	\psi_{\epsilon}(x) = \frac{1}{2}\, \Bigg\{\left[\frac{q_{V,-}}{q_{-}}+\sqrt{\frac{q_{V,+}q_{V,-}}{q_{+}q_{-}}}\right] \, \cos\Big(\sqrt{q_{V_0,+}q_{V_0,-}}a\Big) + \nonumber \\ 
    + i \left[\frac{q_{V,-}}{q_{V_0,-}}\sqrt{\frac{q_{V_0,+}q_{V_0,-}}{q_{+}q_{-}}}
	+\frac{q_{V_0,-}}{q_{-}}\sqrt{\frac{q_{V,+}q_{V,-}}{q_{V_0,+}q_{V_0,-}}}\right] \, \sin\Big(\sqrt{q_{V_0,-}q_{V_0,+}}a\Big)\Bigg\}\begin{pmatrix}
		i\sqrt{q_{-}q_{+}}\\
		q_{-}
	\end{pmatrix}e^{i\sqrt{q_{-}q_{+}}(x-a)} + \nonumber \\
	+\frac{1}{2} \, \Bigg\{\left[\frac{q_{V,-}}{q_{-}}-\sqrt{\frac{q_{V,+}q_{V,-}}{q_{+}q_{-}}}\right] \, \cos\Big(\sqrt{q_{V_0,+}q_{V_0,-}}a\Big) - \nonumber \\
    - i \, \left[\frac{q_{V,-}}{q_{V_0,-}} \, \sqrt{\frac{q_{V_0,-}q_{V_0,+}}{q_{+}q_{-}}}
	 - \frac{q_{V_0,-}}{q_{-}}\sqrt{\frac{q_{V,-}q_{V,+}}{q_{V_0,-}q_{V_0,+}}}\right] \, \sin\Big(\sqrt{q_{V_0,-}q_{V_0,+}}a\Big)\Bigg\}\begin{pmatrix}
		-i\sqrt{q_{-}q_{+}}\\
		q_{-}
	\end{pmatrix}e^{-i\sqrt{q_{-}q_{+}}(x-a)}, \quad \text{for} \quad a<x. \nonumber 
\end{gather}
Where $q_{\pm}=\epsilon\pm m$, $q_{V,\pm}=\epsilon-V\pm m$ and $q_{V_0,\pm}=\epsilon+V_0\pm m$. All these expressions are also applicable for such values of the energy $\epsilon$ that $(\epsilon+V_0)^2-m^2<0$; we just have to use $\sqrt{(\epsilon+V_0)^2-m^2}=i\sqrt{m^2-(\epsilon+V_0)^2}$. Furthermore, the functions $G_1(\epsilon)$, $G_2(\epsilon)$ and $\theta_{\epsilon}$ are defined as:

\begin{gather}
	G_1(\epsilon) = \left|q_{V,-}\right| \, \sqrt{q_{V,+}q_{V,-}} + \frac{\epsilon q_{V,-}}{2}\left(\frac{q_{V,-}}{q_{-}}+\frac{q_{V,+}}{q_{+}}\right) \, \cos^2\Big(\sqrt{q_{V,-}q_{V,+}}a\Big) + \nonumber \\
     + \frac{\epsilon q_{V,-}}{2}\left(\frac{q_{V,-}}{q_{-}}\frac{q_{V_0,+}}{q_{V_0,-}}+\frac{q_{V,+}}{q_{-}}\frac{q_{V_0,-}}{q_{V_0,+}}\right) \, \sin^2\Big(\sqrt{q_{V,+}q_{V,-}}a\Big),
\end{gather}
\begin{gather}
    G_2(\epsilon)=\frac{\epsilon q_{V,-}}{2} \, \left(\frac{q_{V,-}}{q_{-}}-\frac{q_{V,+}}{q_{+}}\right) \, \cos^2\Big(\sqrt{q_{V,+}q_{V,-}}a\Big) + \frac{\epsilon q_{V,-}}{2}\, \left(\frac{q_{V,-}}{q_{+}}\frac{q_{V_0,+}}{q_{V_0,-}}-\frac{q_{V,+}}{q_{+}}\frac{q_{V_0,-}}{q_{V_0,+}}\right) \, \sin^2\Big(\sqrt{q_{V,+}q_{V,-}}a\Big) + \nonumber \\ + i \, \frac{\epsilon q_{V,-}}{2}\, \left(\frac{q_{V_0,-}}{q_{-}}\sqrt{\frac{q_{V,+}q_{V,-}}{q_{V_0,+}q_{V_0,-}}}-\frac{\sqrt{q_{V_0,+}q_{V_0,-}}\sqrt{q_{V,+}q_{V,-}}}{q_{V_0,-}q_{+}}\right) \, \sin\Big(2\sqrt{q_{V_0,+}q_{V_0,-}}a\Big), 
\end{gather}
and

\begin{gather}
\theta_{\epsilon} = \frac{1}{2} \, \arccos\Bigg\{\frac{2G_1(\epsilon)}{(\epsilon-V-m)\sqrt{(\epsilon-V)^2-m^2}}\Big|\frac{q_{V,-}}{q_{-}}\Big|\sqrt{\frac{q_{V,+}q_{V,-}}{q_{+}q_{-}}} \\
\frac{1}{\Big(\frac{q_{V,-}}{q_{-}}-\sqrt{\frac{q_{V,+}q_{V,-}}{q_{+}q_{-}}}\Big)^2\cos^2(\sqrt{q_{V_0,+}q_{V_0,-}}a) + \Big(\frac{q_{V,-}}{q_{V_0,-}}\sqrt{\frac{q_{V_0,-}q_{V_0,+}}{q_{-}q_{+}}}-\frac{q_{V_0,-}}{q_{-}}\sqrt{\frac{q_{V,+}q_{V,-}}{q_{+}q_{-}}}\Big)^2\sin^2(\sqrt{q_{V_0,+}q_{V_0,-}}a)}\Bigg\}.\nonumber 
\end{gather}
We have chosen such a form of the modes because with their use in the nonstationary situation the current can be calculated most straightforwardly. The corresponding calculation is straightforward, but tedious. However, one can find the calculation for a potential of generic form in \cite{Akhmedov:2025jtk}. 

Finally, note that unlike the non-relativistic case, in the relativistic situation one cannot describe the infinite wall case by just taking the limit $V\rightarrow+\infty$ in the expressions above. The problem is that in this limit there are still two modes for the values of energy $|\epsilon|>m$ and both of them are not zero in the region $x<0$ behind the infinite wall. Hence, in this way one cannot obtain the proper boundary conditions for the case of the infinite wall. Below we will use the theory of self-adjoint extensions to find the proper boundary conditions in the case of the infinite wall. The situation with the infinite wall is important for the consideration of supercritical nuclei, in which the radial part of the three-dimensional problem is reduced to the case of the infinite wall (see e.g. \cite{Gitman2012}).

\section{Possible states and resulting expressions for the current}

Having the basis of modes at our disposal, let us describe natural stationary states for the potential in question and calculate the resulting currents. Then we will explain which of the stationary states is realized in the limit $t\to +\infty$ in the non-stationary situation, when the potential is instantly turned on at $t=0$. We start with the case when $V>2m$. We always assume that $V_0 > 2m$.

We define the fermion field operator without any positrons, i.e. as if there are only electronic modes: 

\begin{gather}
	\hat{\psi}(t,x)=\textcolor{black}{\int^{-m}_{-\infty}\sum^2_{j=1}e^{-i\epsilon t}\psi^{(j)}_{\epsilon}(x)\hat{a}^{-}_{\epsilon,j}\frac{d\epsilon}{2\pi}}+\int^{m}_{-m}e^{-i\epsilon t}\psi^{\rightarrow}_{\epsilon}(x)\hat{a}^{-}_{\epsilon,\rightarrow}\frac{d\epsilon}{2\pi}+\int^{V-m}_{m}\sum^2_{j=1}e^{-i\epsilon t}\psi^{(j)}_{\epsilon,\sigma}(x)\hat{a}^{-}_{\epsilon,j}\frac{d\epsilon}{2\pi}+ \nonumber \\
	+\int^{V+m}_{V-m}e^{-i\epsilon t}\psi^{\leftarrow}_{\epsilon}(x)\hat{a}^{-}_{\epsilon,\leftarrow}\frac{d\epsilon}{2\pi}+\int_{V+m}^{+\infty}\sum^2_{j=1}e^{-i\epsilon t}\psi^{(j)}_{\epsilon}(x)\hat{a}^{-}_{\epsilon,j}\frac{d\epsilon}{2\pi}.
\end{gather}
There is, of course, the conjugate operator $\hat{\psi}^{\dagger}(t,x)$ containing the creation operators $\hat{a}^+$ instead of the annihilation ones $\hat{a}^-$.

The second-quantized free Hamiltonian in this case has the following form:

\begin{gather}
	\hat{H} = i \int^{+\infty}_{-\infty}\hat{\psi}^{\dagger}(t,x)\frac{\partial}{\partial t}\hat{\psi}(t,x)dx=\int^{-m}_{-\infty}\epsilon\sum^2_{j=1}\hat{a}^{+}_{\epsilon,j}\hat{a}^{-}_{\epsilon,j}\frac{d\epsilon}{2\pi}+\int^{m}_{-m}\epsilon\hat{a}^{+}_{\epsilon,\rightarrow}\hat{a}^{-}_{\epsilon,\rightarrow}\frac{d\epsilon}{2\pi}+\int^{V-m}_{m}\sum^2_{j=1}\epsilon\hat{a}^{+}_{\epsilon,j}\hat{a}^{-}_{\epsilon,j}+\nonumber \\
	+\int^{V+m}_{V-m}\epsilon\hat{a}^{+}_{\epsilon,\leftarrow}\hat{a}^{-}_{\epsilon,\leftarrow}\frac{d\epsilon}{2\pi}+\int^{+\infty}_{V+m}\epsilon\sum_{j=1}^2\hat{a}^{+}_{\epsilon,j}\hat{a}^{-}_{\epsilon,j}\frac{d\epsilon}{2\pi}.
\end{gather}
The Fock space ground state, the state which is annihilated by all $\hat{a}^-$, we will denote as $|F\rangle$. This state contains the empty Dirac sea:
	\[\hat{a}^{-}_{\epsilon,j}|F\rangle=\hat{a}^{-}_{\epsilon,\leftarrow}|F\rangle=\hat{a}^{-}_{\epsilon,\rightarrow}|F\rangle=0.\]
In fact, then $\hat{\psi}(t,x)|F\rangle=0,$
and this state does not contain any electrons.

But one can also introduce the state that we will denote as $|0\rangle$:

\begin{equation}
    |0\rangle = \prod_{\epsilon<-m}\prod_{j}\hat{a}^{+}_{\epsilon,j}\prod_{-m<\epsilon<0}\hat{a}^{+}_{\epsilon,\rightarrow}|F\rangle.
\end{equation}
This state already contains the Dirac sea and actually has lower energy than $|F\rangle$ because the operator
	
\begin{gather}
    \hat{H} - \langle0|\hat{H}|0\rangle=\int^{-m}_{-\infty}(-\epsilon)\sum_{j=1}^2\hat{a}^{-}_{\epsilon,j}\hat{a}^{+}_{\epsilon,j}\frac{d\epsilon}{2\pi}+\int^{0}_{-m}(-\epsilon)\hat{a}^{-}_{\epsilon,\rightarrow}\hat{a}^{+}_{\epsilon,\rightarrow}\frac{d\epsilon}{2\pi}+\int^{m}_{0}\epsilon\hat{a}^{+}_{\epsilon,\rightarrow}\hat{a}^{-}_{\epsilon,\rightarrow}\frac{d\epsilon}{2\pi}+\nonumber \\ 
	+\int^{V-m}_{m}\epsilon\sum_{j=1}^2\hat{a}^{+}_{\epsilon,j}\hat{a}^{-}_{\epsilon,j}+\int^{V+m}_{V-m}\epsilon\hat{a}^{+}_{\epsilon,\leftarrow}\hat{a}^{-}_{\epsilon,\leftarrow}\frac{d\epsilon}{2\pi}+\int^{+\infty}_{V+m}\epsilon\sum_{j=1}^2\hat{a}^{+}_{\epsilon,j}\hat{a}^{-}_{\epsilon,j}\frac{d\epsilon}{2\pi}
\end{gather}
is positive definite, since all terms in this expression are positive definite. 

The state $|0\rangle$ is the standard Minkowski vacuum or Poincar\'e invariant state. In fact, if one introduces the following operators:

\begin{equation}
    \hat{b}^{-}_{\epsilon,j}=\theta(-m-\epsilon)\hat{a}^{+}_{\epsilon,j},\quad {\rm and} \quad
	\hat{b}^{-}_{\epsilon,\rightarrow}=\theta(m+\epsilon)\theta(-\epsilon)\hat{a}^{+}_{\epsilon,\rightarrow},
\end{equation}
the state $|0\rangle$ can be defined as the Fock space ground state, which is annihilated by new annihilation electron and positron operators\footnote{That follows from the fact that $\hat{a}$ and $\hat{b}$ operators for different values of $\epsilon$ anti-commute with each other and fermionic operators are nilpotent.} $\hat{a}^-$ and $\hat{b}^-$. Note that this is just a renaming of ladder operators rather than an operation that changes the physics. In fact, we just defined our field operator in a different way:
	
\begin{gather}
    \hat{\psi}(t,x)=\int^{+\infty}_{m}\sum_{j=1}^2 e^{i\epsilon t}\psi^{(j)}_{-\epsilon}(x)\hat{b}^{+}_{\epsilon,j}\frac{d\epsilon}{2\pi}+\int^{m}_{0}e^{i\epsilon t}\psi^{\rightarrow}_{-\epsilon}(x)\hat{b}^{+}_{-\epsilon,\rightarrow}\frac{d\epsilon}{2\pi}+\int^{m}_{0}e^{-i\epsilon t}\psi^{\rightarrow}_{\epsilon}(x)\hat{a}^{-}_{\epsilon,\rightarrow}\frac{d\epsilon}{2\pi}+\\
	+\int^{V-m}_{m}\sum_{j=1}^2 e^{-i\epsilon t}\psi^{(j)}_{\epsilon,\sigma}(x)\hat{a}^{-}_{\epsilon,j}\frac{d\epsilon}{2\pi}+\int^{V+m}_{V-m}e^{-i\epsilon t}\psi^{\leftarrow}_{\epsilon}(x)\hat{a}^{-}_{\epsilon,\leftarrow}\frac{d\epsilon}{2\pi}+\int_{V+m}^{+\infty}\sum_{j=1}^2 e^{-i\epsilon t}\psi^{(j)}_{\epsilon}(x)\hat{a}^{-}_{\epsilon,j}\frac{d\epsilon}{2\pi}. \nonumber
\end{gather}
Such a choice of positron ladder operators has a clear physical meaning. In fact, then all particle states have positive energies and the Fock space ground state has the lowest energy. Here we adopt a slightly different way of looking at things and different definitions in comparison with \cite{Akhmedov:2025jtk}. Actually one can perform other types of renaming of some other electron operators to find other states, which are also Fock space ground states. 
 
Now we can find separately contributions to the current of each mode beyond the forbidden zones:

\begin{equation}
	\psi^{(1)*}_{\epsilon}(x)\hat{\alpha}\psi^{(1)}_{\epsilon}(x) = - \psi^{(2)*}_{\epsilon}(x)\hat{\alpha}\psi^{(2)}_{\epsilon}(x) = \cos(2\theta_{\epsilon})\frac{2(\epsilon-V-m)\sqrt{(\epsilon-V)^2-m^2}}{G_1(\epsilon)}.
\end{equation}
The modes from the forbidden zones provide vanishing contributions, as we have explained above.

Finally, we can calculate the current for the above-defined states. In particular, for the lowest energy state $|0\rangle$ the total current is equal to zero: 
\begin{equation}
    \langle0|\hat{j}^{1}(t,x)|0\rangle=\int^{-m}_{-\infty} \Big[\psi^{(1)*}_{\epsilon}(x)\hat{\alpha}\psi^{(1)}_{\epsilon}(x) + \psi^{(2)*}_{\epsilon}(x)\hat{\alpha}\psi^{(2)}_{\epsilon}(x) \Big]\frac{d\epsilon}{2\pi}=0.
\end{equation}
This can be expected on general grounds, because otherwise this state would lose energy, but it has the lowest energy and there is nowhere to lose it. Furthermore, the state $|F\rangle$ also gives zero current, $\langle F|\hat{j}^{1}(t,x)|F\rangle = 0$, simply because $\hat{\psi}(t,x)|F\rangle=0.$

However, one can define yet another state $|K\rangle$\cite{Akhmedov:2025jtk}, which is frequently considered within the context of the Klein phenomenon (see, e.g., the Appendix of \cite{Akhmedov:2020dgc}):

\[|K\rangle=\prod_{V-m>\epsilon>m}\hat{a}^{+}_{\epsilon,1}|0\rangle.\]
It can be seen as a Fock space ground state for another way of defining electron and positron modes within the Klein zone, $\epsilon \in (m,V-m)$. Namely, as we did above in this section, one can rename the corresponding electron annihilation operators as positron creation operators: e.g. $\hat{b}^{-}_{\epsilon,1}=\hat{a}^{+}_{\epsilon,1}$ for $\epsilon \in(m,V-m)$. In any case, the resulting expression for the current in such a state is as follows:

\begin{eqnarray}\label{theresult}
\langle K|\hat{j}^1(t,x)|K\rangle=\int_m^{V-m} \cos(2\theta_{\epsilon})\frac{2(\epsilon-V-m)\sqrt{(\epsilon-V)^2-m^2}}{G_1(\epsilon)}\frac{d\epsilon}{2\pi}.
\end{eqnarray}
The expression for the current practically does not depend on $V_0$ (only via the phase $\theta_\epsilon$). 

As is shown in \cite{Akhmedov:2025jtk} for a similar potential $A_0$ and is explained in the previous section, such a current will appear as the limiting value at $t\to +\infty$ if one instantaneously switches on at $t=0$ the potential $A_0$ under consideration. This happens if as the initial state we choose the Minkowski vacuum. We can show that such a current follows in the non-stationary situation, but adopt here the simple explanation using the stationary considerations.

\subsection{The case when 2m>V}

We continue our considerations with the case when the wall is not sufficiently high, $V<2m$, while the well is deep, i.e. $V_0 > 2m$. Then the analog of the state $|K\rangle$ for the case when $V<2m$ does not contain the Klein zone $\epsilon \in (m,V-m)$ from which the entire contribution to the current was coming, because $V-m < m$ (see eq. (\ref{theresult})).

In the case under consideration the electron field operator is: 

\begin{gather}
\hat{\psi}(t,x) = \int^{-m}_{-\infty}\sum_{j=1}^2e^{-i\epsilon t}\psi_{\epsilon}^{(j)}(x)\hat{a}^{-}_{\epsilon,j}\frac{d\epsilon}{2\pi} +\int^{V-m}_{-m}e^{-i\epsilon t}\psi^{\rightarrow}_{\epsilon}(x)\hat{a}^{-}_{\epsilon,\rightarrow}\frac{d\epsilon}{2\pi}+\sum_{n}e^{-i\epsilon_n t}\psi_{n}(x)\hat{a}^{-}_{n}+\nonumber \\
+\int^{V+m}_{m}e^{-i\epsilon t}\psi^{\leftarrow}_{\epsilon}(x)\hat{a}^{-}_{\epsilon,\leftarrow}\frac{d\epsilon}{2\pi}+\int_{V+m}^{+\infty}\sum_{j=1}^2e^{-i\epsilon t}\psi^{(j)}_{\epsilon}(x)\hat{a}^{-}_{\epsilon,j}\frac{d\epsilon}{2\pi}.
\end{gather}
Similarly to the previous subsection, there are left and right moving modes $\psi^{(1)}_{\epsilon}(x)$ and $\psi^{(2)}_{\epsilon}(x)$ for the range of energies $\epsilon\in(-\infty,-m)\cup(V+m,+\infty)$.
For the range of energies $\epsilon\in(-m;V-m)$ the modes exponentially decay as $x\rightarrow+\infty$, i.e. they are the same as the $\psi^{\rightarrow}_{\epsilon}(x)$ modes from the previous section. Furthermore, there are modes $\psi^{\leftarrow}_{\epsilon}(x)$ for the range $\epsilon\in(m;V+m)$. On top of these modes there is a discrete set of modes $\psi_{n}(x)$ in the region $\epsilon\in(V-m,m)$ with some set of energies $\epsilon_n$ (see the upper Fig.\ref{Fig.1:In_modes_for_big_wall}). This is what one has instead of the Klein zone modes when $V<2m$.

Furthermore, similarly to the previous subsection, we can define the Fock space ground state $|F\rangle$ for the annihilation operators under consideration. One can also define the state $|0\rangle$, which appears to have the lowest energy:

\[|0\rangle=\prod_{\epsilon<-m}\prod_{j}\hat{a}^{+}_{\epsilon,j}\prod_{-m<\epsilon<0}\hat{a}^{+}_{\epsilon,\rightarrow}|F\rangle.\]
For both $|0\rangle$ and $|F\rangle$ the current is zero. But now we essentially cannot define the analog of the state $|K\rangle$ from the previous subsection, because in the range $\epsilon \in (V-m,m)$ there is a discrete set of modes which decay in both directions into the forbidden zones, $x<0$ and $x>a$. Each such mode separately gives a vanishing contribution to the current.

\section{The case of an infinite wall and a well of depth $V_0$}

\begin{figure}[h!]
		\centering
		\includegraphics[width=0.65\linewidth]{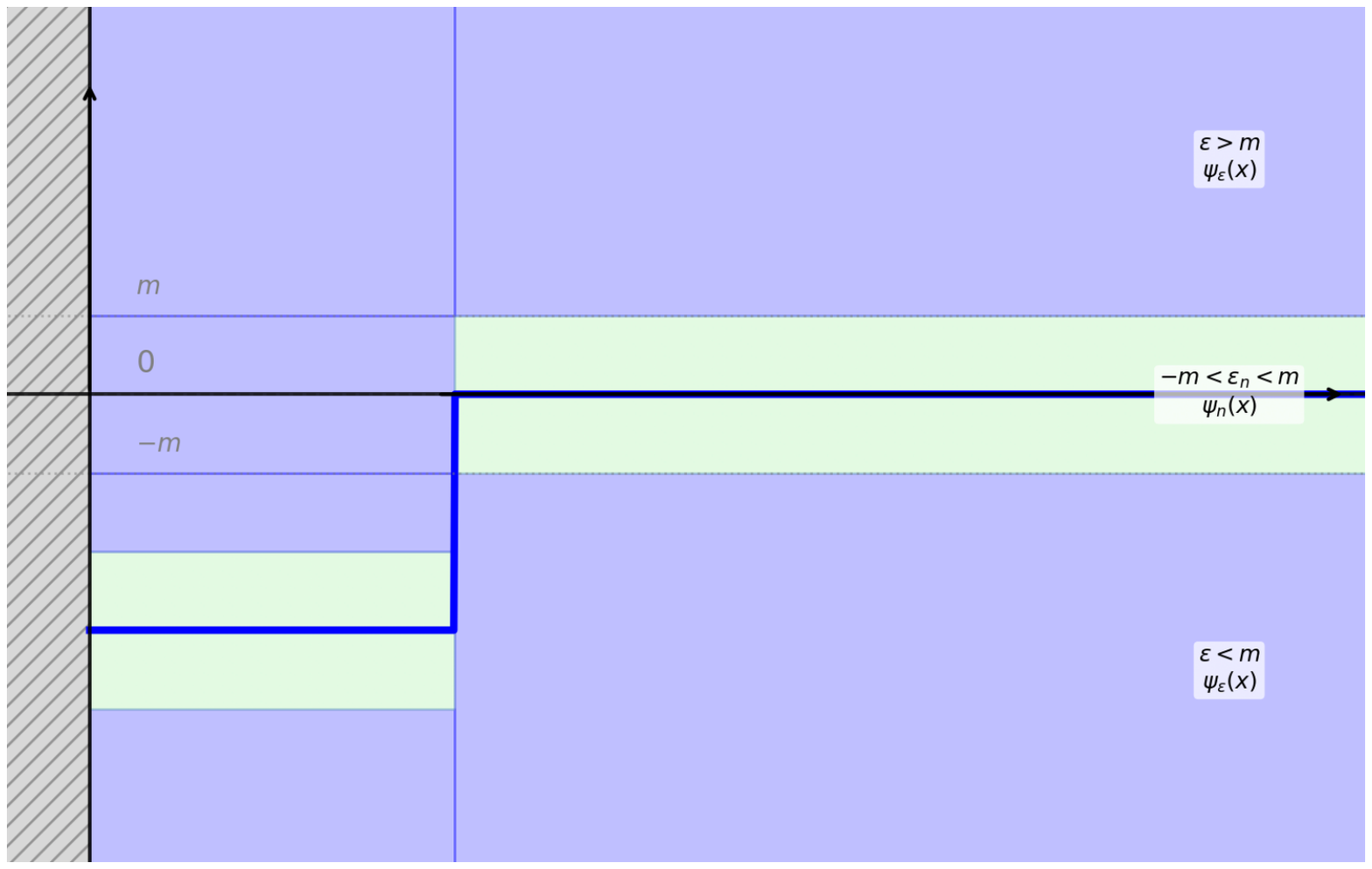}
		\caption{Illustration of the stationary modes for the case of an infinite wall}
		\label{Fig.2:In_modes_for_infinite_wall}
	\end{figure}

The case of an infinite wall and a well of depth $V_0>m$, as depicted in Fig. \ref{Fig.2:In_modes_for_infinite_wall}, is interesting from the perspective that it is related to the decay of supercritical nuclei, once in a spherically symmetric potential the 3D problem is reduced to the 1D radial one.  We discuss here explicitly only the rectangular well, but it is not very hard to extend our observations to the Coulomb potential $\beta/x$ with the appropriate $\beta$ to describe supercritical nuclei, or to the regularized Coulomb potential describing nuclei of finite size.

In non-relativistic quantum mechanics the boundary conditions for the infinite wall can be obtained as the limiting case from the conditions in the potential $U_{V}(x)=\theta(-x)V+\alpha\delta(x)$ in the limit $V\rightarrow+\infty$. Then the conditions will functionally depend on the parameter
$\alpha$. The presence of the delta-functional potential on the half-line $x>0$ affects only the boundary conditions.

However, there is also another way to approach the problem of the definition of the boundary conditions for the infinite wall. The point is that the 1-dimensional non-relativistic Hamiltonian operator with the infinite wall naively is not a self-adjoint operator, because it has imaginary eigenvalues: e.g. any eigenfunction behaving like $e^{-kx}$ is normalizable on the half-line $x>0$. A way to make it self-adjoint is to extend its Hilbert space to include wavefunctions with appropriate boundary conditions at $x=0$. Namely, the extension of the space of functions on which such a Hamiltonian acts is some subspace of the Hilbert space $L^2(R_{+})$, whose definition contains certain boundary conditions. One can obtain these boundary conditions from the general scheme of such extensions (see e.g. \cite{Gitman2012}), but in the particular example under consideration we can classify all such extensions by a parameter $\theta\in R$. One can introduce this parameter $\theta$ in a similar way to the above parameter $\alpha$, because the delta-functional potential in the above example contributes only to the boundary conditions. 
	
Thus, once in the relativistic situation one cannot obtain the boundary conditions for the infinite wall
as the limiting case of a very tall wall, the only way to address the problem is to use the general construction of self-adjoint extensions. We consider the Hamiltonian operator

\begin{equation}
    \hat{H}=i\hat{\alpha}\partial_{x}+\hat{\beta}m-\theta(a-x)V_0
\end{equation}
on the half-line $x\geq 0$. We define the Hilbert space on which it acts as 

\begin{equation}
D_{\theta}(\hat{H})=\left\{\psi(x): \psi(x)\in\begin{pmatrix} L^2(R_{+})\\
		L^2(R_{+}) \end{pmatrix}; \psi(0)=A\begin{pmatrix} 1\\
		\theta \end{pmatrix};\forall A\in C\right\},
\end{equation}
with any parameter $\theta\in R$. We denote the corresponding modes as $\psi^{(\theta)}_{\epsilon}(x)$. Because of these boundary conditions at $x=0$ we obtain single mode for each energy $|\epsilon|>m$, which is a linear combination of left and right moving waves, rather than two (separately left or right moving) modes, as was the case for the tall wall considered in the previous sections. 
	
Now the electron field operator is as follows:

\begin{equation}
    \hat{\psi}(t,x)=\int^{-m}_{-\infty}e^{-i\epsilon t}\psi_{\epsilon}^{(\theta)}(x)\hat{a}^{-}_{\epsilon}\frac{d\epsilon}{2\pi}+\sum_{n}e^{-i\epsilon_n t}\psi_{n}^{(\theta)}(x)\hat{a}^{-}_{n}+\int^{+\infty}_{m}e^{-i\epsilon t}\psi^{(\theta)}_{\epsilon}(x)\hat{a}^{-}_{\epsilon}\frac{d\epsilon}{2\pi}.
\end{equation}
The discrete levels $\psi_{n}(x)$ are present for the range of energies $-m<\epsilon_n<m$. They exponentially decay into the forbidden zone, as shown in Fig. \ref{Fig.2:In_modes_for_infinite_wall}.

The form of the modes from the continuous part of the spectrum, i.e. for $|\epsilon|>m$, is as follows:

\begin{eqnarray}
    \psi^{(\theta)}_{\epsilon}(x)=\Big(\sqrt{\frac{q_{-}}{q_{+}}}\Big[\frac{\sqrt{q_{V_0,+}q_{V_0,-}}}{q_{V_0,-}}\cos(\sqrt{q_{V_0,+}q_{V_0,-}}a)-\frac{q_{V_0,+}}{q_{V_0,-}}\theta \sin(\sqrt{q_{V_0,+}q_{V_0,-}}a)\Big]^2+\nonumber \\
	+\sqrt{\frac{q_{+}}{q_{-}}}\Big[\frac{\sqrt{q_{V_0,+}q_{V_0,-}}}{q_{V_0,-}}\theta \cos(\sqrt{q_{V_0,+}q_{V_0,-}}a)+\sin(\sqrt{q_{V_0,+}q_{V_0,-}}a)\Big]^2\Big)^{-\frac{1}{2}}\times \nonumber \\
	\begin{pmatrix}
		\frac{\sqrt{q_{V_0,+}q_{V_0,-}}}{q_{V_0,-}}\cos(\sqrt{q_{V_0,+}q_{V_0,-}}x)-\frac{q_{V_0,+}}{q_{V_0,-}}\theta \sin(\sqrt{q_{V_0,+}q_{V_0,-}}x)\\
		\frac{\sqrt{q_{V_0,+}q_{V_0,-}}}{q_{V_0,-}}\theta \cos(\sqrt{q_{V_0,+}q_{V_0,-}}x)+\sin(\sqrt{q_{V_0,+}q_{V_0,-}}x)
	\end{pmatrix}, \quad \text{for } \quad x<a; \\
\psi^{(\theta)}_{\epsilon}(x)=\frac{\sqrt{q_{+}}}{\sqrt[4]{q_{+}q_{-}}}\begin{pmatrix}
		\cos(\sqrt{q_{+}q_{-}}x+\delta(\epsilon))\\
		\frac{\sqrt{q_{+}q_{-}}}{q_{+}}\sin(\sqrt{q_{+}q_{-}}x+\delta(\epsilon))
	\end{pmatrix},\quad \text{for } \quad x>a, \nonumber 
\end{eqnarray}
where
	\[\delta(\epsilon)=\arctan\Big(\text{sgn}(\epsilon)\sqrt{\frac{q_{+}}{q_{-}}}\frac{\sqrt{q_{V_0,+}q_{V_0,-}}\cos(\sqrt{q_{V_0,+}q_{V_0,-}}a)-q_{V_0,+}\theta \sin(\sqrt{q_{V_0,+}q_{V_0,-}}a)}{\sqrt{q_{V_0,+}q_{V_0,-}}\theta \cos(\sqrt{q_{V_0,+}q_{V_0,-}}a)+q_{V_0,-}\sin(\sqrt{q_{V_0,+}q_{V_0,-}}a)}\Big)-a\sqrt{q_{+}q_{-}}\]
is some phase. These modes together form an orthonormal and complete basis.


As in previous sections we can define the Fock space ground state $|F\rangle$ with the empty Dirac sea and also the state $|0\rangle$ with the lowest energy:

\begin{equation}
|0\rangle=\prod_{\epsilon<-m}\hat{a}^{+}_{\epsilon}\prod_{n:\epsilon_n<0}\hat{a}^{+}_{n}|F\rangle
\end{equation}
Furthermore, we can introduce positron operators:

\begin{eqnarray}
    \hat{b}^{-}_{\epsilon}=\theta(-m-\epsilon)\hat{a}^{+}_{\epsilon}, \quad \hat{b}^{-}_{n}=\theta(-\epsilon_n)\hat{a}^{+}_{n},
\end{eqnarray}
to define the state $|0\rangle$ as the Fock space ground state for the electron and positron annihilation operators. 

However, similarly to the situation in section 4.1 one cannot introduce the state $|K\rangle$. That is true essentially because there is no Klein zone in the potential under consideration.  In fact, due to the boundary conditions all modes are linear combinations of left- and right-moving waves and, hence:

\[\psi^{(\theta)*}_{\epsilon}(x)\hat{\alpha}\psi^{(\theta)}_{\epsilon}(x)=0, \quad \psi^{(\theta)*}_{n}(x)\hat{\alpha}\psi^{(\theta)}_{n}(x)=0.\]
Thus, for any stationary density matrix the expectation value of the current is zero:

	\[{\rm Tr}\left[\hat{\rho}\hat{\psi}^{\dagger}(t,x)\hat{\alpha}\hat{\psi}(t,x)\right] =\int_{|\epsilon|>m}{\rm Tr}\left(\hat{\rho}\hat{a}^{+}_{\epsilon}\hat{a}^{-}_{\epsilon}\right) \, \left[\psi^{(\theta)*}_{\epsilon}(x)\hat{\alpha}\psi^{(\theta)}_{\epsilon}(x) \right] \, \frac{d\epsilon}{2\pi} + \sum_{n} {\rm Tr} \left(\hat{\rho}\hat{a}^{+}_{n}\hat{a}^{-}_{n}\right) \, \left[\psi^{(\theta)*}_{n}(x)\hat{\alpha}\psi^{(\theta)}_{n}(x)\right] = 0.\]
Furthermore, if we instantaneously create a well of depth $V_0>2m$  in the presence of an infinite wall, no current will be created in the Gaussian approximation, because in such a case one cannot create a state, in which only left or right moving states are filled in.

\section{Conclusions and acknowledgments}

Thus, we observe that two parallel capacitors with oppositely directed strong electric fields create pairs only if the potential energy difference between them is large enough, $V>2m$. We find this observation quite surprising. Of course, we make these observations in $1+1$ dimensional theory and in the Gaussian approximation. Hence, one has to check whether the situation does not substantially change in $3+1$ QED. However, it is hard to believe that the presence of transverse momenta in the mode functions will drastically change the physics of the phenomenon in question.  

Furthermore, we observe that the presence of an infinite wall, i.e. the problem on the half-line $x>0$ with certain boundary conditions, prevents the creation of particles essentially for any well in the Gaussian approximation. We show this explicitly for the rectangular well, but it is not very hard to extend our observations to the Coulomb potential $\beta/x$ with the appropriate $\beta$ to describe supercritical nuclei, or to the regularized Coulomb potential describing nuclei of finite size. Actually, our observations are in accordance with the classical literature \cite{landau1983quantum} (see also \cite{gershtein1970positron}), which states that supercritical nuclei will create simultaneously a couple of electron-positron pairs rather than a single pair. To describe such a process, one has to go beyond the Gaussian approximation.

We would like to acknowledge discussions with D. Sadekov, P. Zabgorodny, K. Kazarnovskii and D. Diakonov. The work was performed under the financial support of the RFBR grant 26-72-10125.

	\bibliography{literature}

@article{Schwinger:1951nm,
    author = "Schwinger, Julian S.",
    editor = "Milton, K. A.",
    title = "{On gauge invariance and vacuum polarization}",
    doi = "10.1103/PhysRev.82.664",
    journal = "Phys. Rev.",
    volume = "82",
    pages = "664--679",
    year = "1951"
}

@article{Akhmedov:2025jtk,
    author = "Akhmedov, E. T. and Diakonov, D. V. and Lapushkin, V. I. and Sadekov, D. I.",
    title = "{Lessons from the Klein paradox}",
    eprint = "2512.24770",
    archivePrefix = "arXiv",
    primaryClass = "hep-th",
    doi = "10.1103/1mb6-7hb4",
    journal = "Phys. Rev. D",
    volume = "113",
    number = "8",
    pages = "085016",
    year = "2026"
}

@article{Gavrilov:2015yha,
    author = "Gavrilov, S. P. and Gitman, D. M.",
    title = "{Quantization of charged fields in the presence of critical potential steps}",
    eprint = "1506.01156",
    archivePrefix = "arXiv",
    primaryClass = "hep-th",
    doi = "10.1103/PhysRevD.93.045002",
    journal = "Phys. Rev. D",
    volume = "93",
    number = "4",
    pages = "045002",
    year = "2016"
}

@article{Calogeracos:1998rf,
    author = "Calogeracos, A. and Dombey, Norman",
    title = "{Klein tunneling and the Klein paradox}",
    eprint = "quant-ph/9806052",
    archivePrefix = "arXiv",
    reportNumber = "SUSX-TH-97-019, SUSX-TH-97-019",
    doi = "10.1142/S0217751X99000312",
    journal = "Int. J. Mod. Phys. A",
    volume = "14",
    pages = "631--644",
    year = "1999"
}

@article{ternov2024paradoks15483,
author = {Ternov, A. I.},
title = {The Klein Paradox and Klein Tunneling},
journal = {Memoirs of the Faculty of Physics},
year = 2024,
number = 5,
language = english
}

@article{Akhmedov:2020dgc,
    author = "Akhmedov, E. T. and Anokhin, A. V. and Sadekov, D. I.",
    title = "{Currents of created pairs in strong electric fields}",
    eprint = "2012.00399",
    archivePrefix = "arXiv",
    primaryClass = "hep-th",
    doi = "10.1142/S0217751X21501347",
    journal = "Int. J. Mod. Phys. A",
    volume = "36",
    number = "19",
    pages = "2150134",
    year = "2021"
}

@article{Nikishov1985,
  author = {Nikishov, A. I.},
  title = {Problems of intense external-field intensity in quantum electrodynamics},
  journal = {Journal of Soviet Laser Research},
  volume = {6},
  number = {6},
  pages = {619--717},
  year = {1985},
  month = {11},
  issn = {1573-8760},
  doi = {10.1007/BF01120143},
  url = {https://doi.org/10.1007/BF01120143}
}

@article{Nikishov:1970br,
    author = "Nikishov, A. I.",
    title = "{Barrier scattering in field theory removal of klein paradox}",
    doi = "10.1016/0550-3213(70)90527-4",
    journal = "Nucl. Phys. B",
    volume = "21",
    pages = "346--358",
    year = "1970"
}

@article{Akhmedov:2021rhq,
    author = "Akhmedov, E. T.",
    title = "{Curved space equilibration versus flat space thermalization: A short review}",
    eprint = "2105.05039",
    archivePrefix = "arXiv",
    primaryClass = "gr-qc",
    doi = "10.1142/S0217732321300202",
    journal = "Mod. Phys. Lett. A",
    volume = "36",
    number = "20",
    pages = "2130020",
    year = "2021"
}

@article{Calogeracos:1999yp,
    author = "Calogeracos, A. and Dombey, Norman",
    title = "{History and physics of the Klein paradox}",
    eprint = "quant-ph/9905076",
    archivePrefix = "arXiv",
    reportNumber = "SUSX-TH-99-032, SUSX-TH-99-032",
    doi = "10.1080/001075199181387",
    journal = "Contemp. Phys.",
    volume = "40",
    pages = "313--321",
    year = "1999"
}

@article{Klein,
    author = "O. Klein",
    title = "{Die Reflexion von Elektronen an einem Potentialsprung nach der relativistischen Dynamik von Dirac}",
    doi = "doi:10.1007/BF01339716",
    journal = "Zeitschrift für Physik",
    volume = "53",
    pages = "157--165",
    year = "1929"
}

@article{landau1983quantum,
  title={Quantum electrodynamics},
  author={Landau, Lev Davidovich and Lifshitz, EM},
  year={1983},
  publisher={Pergamon Pr}
}

@article{gershtein1970positron,
  title={Positron production during the mutual approach of heavy nuclei and the polarization of the vacuum},
  author={S.S.Gershtein and Ya.B.Zeldovich},
  journal={Sov. Phys. JETP},
  volume={30},
  number={2},
  pages={358--361},
  year={1970}
}

@article{Akhmedov:2026wew,
    author = "Akhmedov, E. T. and Zavgorodny, P. S.",
    title = "{Does Schwinger{\textquoteright}s value for the current of created pairs get modified for a long and strong enough pulse?}",
    eprint = "2601.12847",
    archivePrefix = "arXiv",
    primaryClass = "hep-th",
    doi = "10.1103/4x7v-447k",
    journal = "Phys. Rev. D",
    volume = "113",
    number = "10",
    pages = "105012",
    year = "2026"
}

@article{Akhmedov:2024npw,
    author = "Akhmedov, E. T. and Lapushkin, V. I. and Sadekov, D. I.",
    title = "{Light fields in various patches of de~Sitter spacetime}",
    eprint = "2411.11106",
    archivePrefix = "arXiv",
    primaryClass = "hep-th",
    doi = "10.1103/k65j-1jn4",
    journal = "Phys. Rev. D",
    volume = "111",
    number = "12",
    pages = "125015",
    year = "2025"
}

@article{Akhmedov:2024rkt,
    author = "Akhmedov, E. T. and Zavgorodny, P. S.",
    title = "{Higher loop corrections to the current of created pairs in the lengthy electric pulse}",
    eprint = "2409.00684",
    archivePrefix = "arXiv",
    primaryClass = "hep-th",
    doi = "10.1103/r6y3-bdql",
    journal = "Phys. Rev. D",
    volume = "111",
    number = "12",
    pages = "125017",
    year = "2025"
}

@article{Akhmedov:2024lce,
    author = "Akhmedov, E. T. and Anokhin, A. V. and Kazarnovskii, K. A.",
    title = "{Relevance of quantum corrections to the matter stress-energy tensor in eternally expanding universes}",
    eprint = "2401.12855",
    archivePrefix = "arXiv",
    primaryClass = "hep-th",
    doi = "10.1103/PhysRevD.109.085012",
    journal = "Phys. Rev. D",
    volume = "109",
    number = "8",
    pages = "085012",
    year = "2024"
}

@article{Akhmedov:2023zfy,
    author = "Akhmedov, E. T. and Zavgorodny, P. S. and Sadekov, D. I. and Kazarnovskii, K. A.",
    title = "{Loop corrections to the current of pairs created in a lengthy electric pulse}",
    eprint = "2303.08624",
    archivePrefix = "arXiv",
    primaryClass = "hep-th",
    doi = "10.1103/PhysRevD.107.125006",
    journal = "Phys. Rev. D",
    volume = "107",
    number = "12",
    pages = "125006",
    year = "2023"
}

@book{Gitman2012,
    author    = {Gitman, D. M. and Tyutin, I. V. and Voronov, B. L.},
    title     = {Self-adjoint Extensions in Quantum Mechanics: General Theory and Applications to Schr{\"o}dinger and Dirac Equations with Singular Potentials},
    publisher = {Birkhäuser},
    year      = {2012},
    series    = {Progress in Mathematical Physics},
    volume    = {62},
    doi       = {10.1007/978-0-8176-4662-2},
    isbn      = {978-0-8176-4400-0}
}
	\bibliographystyle{unsrt}

\end{document}